\documentclass{elsart}
\usepackage{latexsym}
\usepackage{epsfig}
\usepackage{natbib}

\def\url#1{#1} % temporally

\begin{document}

\begin{frontmatter}

\title{A Simplified Model of ADAF with the Jet Driven by the Large-Scale
Magnetic Field}

%\pubyear{Submitted}
%\pagerange{\pageref{firstpage}--\pageref{endpage}}
%\onecolumn
\author {Yang Li, Zhao-Ming Gan and Ding-Xiong Wang$^{1}$}
\address{School of Physics, Huazhong University of Science
and Technology, Wuhan 430074, China }

      \thanks[email]{Corresponding author. Tel: +862787540742; fax:
      +862787542391\\
      E-mail address: dxwang@mail.hust.edu.cn (D.-X. Wang).}

\begin{abstract}

We propose a simplified model of outflow/jet driven by the Blandford-Payne
(BP) process from advection dominated accretion flows (ADAF) and derive the
expressions of the BP power and disk luminosity based on the conservation
laws of mass, angular momentum and energy. We fit the 2--10 keV luminosity
and kinetic power of 15 active galactic nucleus (AGNs) of sub-Eddington
luminosity. It is found that there exists an anti-correlation between the
accretion rate and the advection parameter, which could be used to explain
the correlation between Eddington-scaled kinetic power and bolometric
luminosity of the 15 samples. In addition, the Ledlow-Owen relation for FR
I/II dichotomy is re-expressed in a parameter space consisting of logarithm
of dimensionless accretion rate versus that of the BH mass. It turns out
that the FR I/II dichotomy is determined mainly by the dimensionless
accretion rate, being insensitive to the BH mass. And the dividing accretion
rate is less than the critical accretion rate for ADAFs, suggesting that FR
I sources are all in the ADAF state.

\end{abstract}


\end{frontmatter}

%-----------------------sec 1 ----------------------------------------

\section{INTRODUCTION}

Advection-dominated accretion flow (ADAF) is widely regarded as a successful
model for explaining the quiescent and hard states of black hole (BH) X-ray
binaries as well as low-luminosity AGNs (Narayan 2005, Yuan 2007, and Ho
2008 for reviews). It have been predicted theoretically and confirmed in
numerical simulations that strong winds and jets can be driven by the
tremendous thermal energy in ADAF (see Narayan {\&} McClintock 2008 and
references therein). Observational evidence for the association of
nonthermal relativistic jets with ADAFs has been accumulated in recent years
with the discovery of radio emission in virtually every BH binaries in the
hard state and in low-luminosity AGNs (Corbel et al. 2000; Fender 2001;
Fender, Belloni {\&} Gallo 2004; Fender {\&} Belloni 2004).

\quad\quad It has been pointed out that an outflow emanating from an
accretion disk can act as a sink for mass, angular momentum and
energy, altering the dissipation rates and effective temperatures
across the disk (Donea {\&} Biermann 1996; Knigge 1999; Kuncic {\&}
Bicknell 2007; Li et al. 2008; Xie {\&} Yuan 2008). Thus the jets
launched from the ADAFs should influence the dynamics of ADAFs due
to the mass, angular momentum and energy extracted through outflow.

\quad\quad The currently most favored mechanisms for jet production
include the Blandford-Znajek (BZ) process (Blandford {\&} Znajek
1977) and the Blandford-Payne (BP) process (Blandford {\&} Payne
1982, hereafter BP82). In the BZ process, energy and angular
momentum are extracted from a rotating BH and transferred to a
remote astrophysical load by open magnetic field lines. In the BP
process, the magnetic fields threading the disk extract energy from
the accretion disk itself to power the jet/outflow. Some authors
argued that the maximal jet power extracted via the BP process
dominates over the power extracted via the BZ process, provided that
the poloidal field threading the BH is not significantly greater
than that threading the inner disk (e.g., Livio et al. 1999; Meier
1999, 2001; Cao 2002; Nemmen et al. 2007). Young et al. (2007)
discovered the large rotating wind of the quasar PG 1700+518,
providing direct observational evidence that outflows from AGNs is
launched from the disks.

\quad\quad Merloni {\&} Heinz (2007, hereafter MH07) obtained a
clear relationship between Eddington-scaled kinetic power and
bolometric luminosity based on a statistical analysis of a sample of
15 sub-Eddington AGNs. The measured slope suggests that these
objects are in a radiatively inefficient accretion mode. And they
interpreted this fact with a simple coupled accretion-outflow disk
model. However, dynamically important magnetic fields and their role
in the jet production mechanisms were not taken into account in
MH07.

\quad\quad Fanaroff {\&} Riley (1974) recognized that the majority
of radio galaxies can be classified into two types (FR I and FR II)
according to their radio morphology, i.e., edge darkened and edge
brightened sources, and that this division rather neatly translates
into a separation in radio power being below and above $L_{178}
\simeq 2.5\times 10^{33}h_{50}^{ - 2} \mbox{ erg s}^{ - 1}\mbox{
Hz}^{ - 1}$ at 178 MHz. This division has become even clearer and
sharper when it has been found by Ledlow {\&} Owen (1994, 1996) that
it is clearly shown by a line in the plane of the optical luminosity
of the host galaxy and the total radio luminosity. Over the years,
much work has been done to understand the remarkable FR I/II
dichotomy, which may depend on the fundamental parameters, such as
BH spin and accretion mode (Meier, 1999; Ghisellini {\&} Celotti,
2001, hereafter GC01); Wold, Lacy {\&} Armus 2007, hereafter WLA07;
Wu {\&} Cao 2008, hereafter WC08).

\quad\quad Motivated by the above work, we intend to discuss the
outflow/jet driven by the BP process, and investigate the
interaction of the outflow/jet with ADAF. For simplicity we treat a
Schwarzschild BH, and the BZ process is not included in this model.
Following Yuan, Ma {\&} Narayan (2008, hereafter YMN08), we replace
the radial momentum equation by a simple algebraic relation between
the angular velocity of the gas and the Keplerian angular velocity
of the disk. Based on the conservation laws of mass, angular
momentum and energy we derive the expressions of the BP power and
disk luminosity. We fit the 2--10 keV luminosity and kinetic power
of the 15 samples given in MH07 based on ADAF model with jet, and
obtain an anti-correlation between the accretion rate and the
advection parameter, which could be used to explain the correlation
between Eddington-scaled kinetic power and bolometric luminosity of
the 15 samples. In addition, we express the dividing line of the
Ledlow-Owen relation for FR I/II dichotomy in the plane of $\lg \dot
{m}_H $ versus $\lg M_{BH} $. We find that the FR I/II dichotomy is
closely related to the accretion rate, but has a weak dependence on
the BH mass, which is consistent with the results derived by GC01
and WLA07. And the dividing accretion rate is less than the critical
accretion rate for ADAFs, suggesting that FR I sources is in the
ADAF state.

\quad\quad This paper is organized as follows. In Sect. 2 we
describe our model, and derive the expressions for the BP power and
disk luminosity at the presence of a jet based on the conservation
laws of mass, angular momentum and energy. In Sect. 3, we fit the
2--10 keV luminosity and kinetic power of the 15 sources given in
MH07 based on our model. In Sect. 4 we present the dividing lines of
the Ledlow-Owen relation for FR I/II dichotomy in the plane of $\lg
\dot {m}_H - \lg M_{BH} $. Finally, in Sect. 5, we summarize our
main conclusions.

%-----------------------sec 2 ----------------------------------------

\section{MODEL DESCRIPTION}

We assume that the ADAF is stationary and axisymmetric, extending from the
outer edge to the BH horizon. Two kinds of magnetic fields are involved in
this model, i.e., the large-scale magnetic field threading the ADAF and the
small-scale magnetic field tangled in the ADAF. The large-scale and the
small-scale magnetic fields are assumed to work independently, contribute to
the BP process and viscosity, respectively. The ADAF is assumed to be
ideally conducting and force-free.

\quad\quad Following BP82, we assume that the poloidal magnetic
field on the disk surface varies with the disk radius as follows,

\begin{equation}
\label{eq1}
B_{ADAF} = B_H \left( {r \mathord{\left/ {\vphantom {r {r_H }}} \right.
\kern-\nulldelimiterspace} {r_H }} \right)^{ - 5 \mathord{\left/ {\vphantom
{5 4}} \right. \kern-\nulldelimiterspace} 4},
\end{equation}

\noindent
where $r$ and $r_H $ are the ADAF and horizon radii, respectively, and we
have $r_H = 2r_g $ with $r_g = {GM} \mathord{\left/ {\vphantom {{GM} {c^2}}}
\right. \kern-\nulldelimiterspace} {c^2}$ for a Schwarzschild BH. The
quantities $B_{ADAF} $ and $B_H $ are the magnetic field at ADAF and
horizon, respectively.

\quad\quad Considering the balance between the magnetic pressure on
the horizon and the ram pressure of the innermost parts of ADAF,
Moderski, Sikora {\&} Lasota (1997) expressed the magnetic field at
the horizon as follows,

\begin{equation}
\label{eq2}
{B_H^2 c} \mathord{\left/ {\vphantom {{B_H^2 c} {8\pi }}} \right.
\kern-\nulldelimiterspace} {8\pi } = P_{ram} \sim \rho c^2\sim {\dot {M}_H
c^2} \mathord{\left/ {\vphantom {{\dot {M}_H c^2} {\left( {4\pi r_H^2 }
\right)}}} \right. \kern-\nulldelimiterspace} {\left( {4\pi r_H^2 }
\right)},
\end{equation}

\noindent
where $\dot {M}_H $ is the accretion rate at the BH horizon, which can be
written as a dimensionless one in terms of the Eddington accretion rate as
follows,

\begin{equation}
\label{eq3}
\dot {m}_H = {\dot {M}_H } \mathord{\left/ {\vphantom {{\dot {M}_H } {\dot
{M}_{Edd} }}} \right. \kern-\nulldelimiterspace} {\dot {M}_{Edd} },
\end{equation}

\noindent
where

\begin{equation}
\label{eq4}
\dot {M}_{Edd} = 1.4\times 10^{18}M_{BH} (M_ \odot )\left( {g \cdot s^{ -
1}} \right).
\end{equation}

\quad\quad Considering that equation (\ref{eq2}) is uncertain, we
introduce a parameter $\lambda $ to adjust the magnetic pressure at
the BH horizon relative to the ram pressure of ADAF, and equation
(\ref{eq2}) is rewritten as

\begin{equation}
\label{eq5}
\dot {M}_H c = \lambda B_H^2 r_H^2 .
\end{equation}

\quad\quad As argued by BP82 and Spruit (1996), the outflow matter
could be accelerated centrifugally along the magnetic field lines,
overcoming a barrier of gravitational potential to form
magnetohydrodynamic (MHD) jets. The Poynting flux dominates over the
kinetic flux near the disk surface, and the former is converted into
the latter during accelerating matter in the outflow. Thus we
express the Poynting energy flux $S_E $ extracted
electromagnetically from the ADAF as (The derivation of equation
(\ref{eq6}) is given in Appendix A),

\begin{equation}
\label{eq6}
S_E = \frac{B_{ADAF}^2 \Omega ^2r^2}{4\pi c}.
\end{equation}

\quad\quad The electromagnetic angular momentum flux $S_L $
extracted is related to the Poynting energy flux by

\begin{equation}
\label{eq7}
S_L = {S_E } \mathord{\left/ {\vphantom {{S_E } \Omega }} \right.
\kern-\nulldelimiterspace} \Omega = \frac{B_{ADAF}^2 \Omega r^2}{4\pi c}.
\end{equation}

The energy flux of the jet can be regarded as the sum of the kinetic and the
Poynting fluxes,

\begin{equation}
\label{eq8}
F_{jet} = \frac{1}{2}\dot {M}_{jet} \left( {\Omega ^2r^2 + \upsilon _P^2 }
\right) + S_E = F_K + S_E ,
\end{equation}

\noindent
where $\dot {M}_{jet} $ is the mass loss rate in the jet from unit area of
the ADAF, and $\upsilon _P $ is the poloidal velocity of the outflow, and
$F_K $ is the kinetic flux.

\quad\quad It is noted that $F_K $ is much less than $S_E $ at the
surface of ADAF, while the two are comparable at Alfven surface due
to a considerable fraction of the Poynting flux converted into
kinetic flux. Thus the energy flux of the jet is related to $\dot
{M}_{jet} $ by

\begin{equation}
\label{eq9}
F_{jet} = \Gamma _j \dot {M}_{jet} c^2,
\end{equation}

\noindent
where $\Gamma _j $ is the bulk Lorentz factor of the jet.

\quad\quad Considering that $F_{jet} $ is dominated by $S_E $ at the
surface of ADAF, and combining equations (\ref{eq1}), (\ref{eq5}),
(\ref{eq6}), (\ref{eq8}) and (\ref{eq9}), we derive the relation
between $\dot {M}_{jet} $ and $\dot {M}_H $ as follows,

\begin{equation}
\label{eq10}
\dot {M}_{jet} = {\dot {M}_H r_H^{0.5} r^{ - 0.5}\Omega ^2} \mathord{\left/
{\vphantom {{\dot {M}_H r_H^{0.5} r^{ - 0.5}\Omega ^2} {\left( {4\pi \lambda
\Gamma _j c^2} \right)}}} \right. \kern-\nulldelimiterspace} {\left( {4\pi
\lambda \Gamma _j c^2} \right)}.
\end{equation}

Inspecting equation (\ref{eq10}), we find that the ratio of $\dot {M}_{jet} $ to
$\dot {M}_H $ is sensitive to and inversely proportional to the value of the
parameter $\lambda $.

\quad\quad Based on the self-similar solution obtained by Narayan
{\&} Yi (1994) YMN08 suggested a simple algebraic relation between
the angular velocity of ADAF and the Keplerian angular velocity as
follows,

\begin{equation}
\label{eq11}
\Omega = f\Omega _K ,
\end{equation}

\noindent
where the quantity $f$ is a function of the radius $r$ expressed as follows,

\begin{equation}
\label{eq12} f = \left\{ {\begin{array}{l}
 f_0 \mbox{, \ \ \ \ \ \ \ \  \ \ \ \ \ \ \ \ \ \ \ for }r \ge 6r_g , \\
 f_0 {3\left( {r - 2r_g } \right)} \mathord{\left/ {\vphantom {{3\left( {r -
2r_g } \right)} {2r}}} \right. \kern-\nulldelimiterspace}
{2r}\mbox{, \ for }r
< 6r_g \mbox{, } \\
 \end{array}} \right.
\end{equation}

In equation (\ref{eq12}) $f_0 $ is an adjustable constant. As argued in YMN08, $f_0
= 0.33$ gives very good results for all accretion rates as the viscous
parameter $\alpha $ is large, say $ \sim 0.3$, for which an ADAF solution is
possible. Although the global solutions of ADAF with large-scale magnetic
field have not been achieved, we think equations (\ref{eq11}) and (\ref{eq12}) could be
applicable to ADAF with jet driven by the BP process. We expect that the
values of the function $f$ are less than those given by equation (\ref{eq12}), and $f_0
= 0.33$ can be regarded as an upper limit in calculations.

\quad\quad Incorporating equations (\ref{eq10}), (\ref{eq11}) and
(\ref{eq12}), we have the dimensionless mass flux of the jet as
follows,

\begin{equation}
\label{eq13}
\begin{array}{l}
\dot {m}_{jet} (\lambda ,\Gamma _j ,\xi ) = \dot {M}_{jet} \left(
{{4\pi r_g^2 } \mathord{\left/ {\vphantom {{4\pi r_g^2 } {\dot {M}_H
}}}
\right. \kern-\nulldelimiterspace} {\dot {M}_H }} \right) \\
\mbox{ }\mbox{ }\mbox{ }\mbox{ }\mbox{ }\mbox{ }\mbox{ }\mbox{
}\mbox{ }\mbox{ }\mbox{ }\mbox{ }\mbox{ }\mbox{ }=\frac{2f_0^2 \xi
_H^{0.5} }{\lambda \Gamma _j }\xi ^{ - 3.5}\times \left\{
{\begin{array}{l}
 \mbox{1, \ \ \ \ \ \ \ \ \ \ \ \ \ \ \ \ \ \ for }\xi \ge 6, \\
 \left[ {{3\left( {\xi - 2} \right)} \mathord{\left/ {\vphantom
{{3\left( {\xi - 2} \right)} {2\xi }}} \right.
\kern-\nulldelimiterspace} {2\xi }}
\right]^{2} \mbox{, \ for }  \xi < 6, \\
 \end{array}} \right.\\
 \end{array}
\end{equation}

\noindent
where $\xi \equiv r \mathord{\left/ {\vphantom {r {r_g }}} \right.
\kern-\nulldelimiterspace} {r_g }$ is the radial parameter of the disk
defined in terms of the radius $r_g $.

\quad\quad Based on equation (\ref{eq13}) we have the curves of the
dimensionless jet mass flux $\dot {m}_{jet} $ varying with the
parameter $\xi $ for the given values of $\Gamma _j $ as shown in
Figure 1.

\quad\quad As shown in Figure 1, the dimensionless jet mass flux
$\dot {m}_{jet} $ increases with $\xi $ at first, and then it
decreases steeply with $\xi $ as $\dot {m}_{jet} \propto \xi ^{ -
3.5}$, attaining its peak value in the region $r < 4r_g $. This
result implies that the outflow is launched predominantly from the
innermost ADAF.

\quad\quad Inspecting equation (\ref{eq13}) and Figure 1, we find
that the dimensionless mass flux of the jet at the given radius
decreases significantly with the increasing $\Gamma _j $, and we
have $\dot {m}_{jet} \to 0$ for $\Gamma _j \to \infty $. This result
implies that the extracted energy from the disk is almost carried in
the form of electromagnetic form, suggesting the presence of a
strong mass loss in the case that the bulk Lorentz factor is small
enough.

%-----------------------fig 1 ----------------------------------------

\begin{figure}
\vspace{0.5cm}
\begin{center}
\includegraphics[width=7cm]{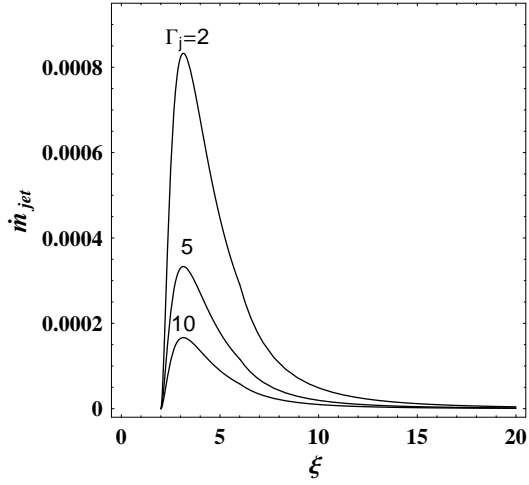}
\caption{ The curves of $\dot {m}_{jet} $ versus $\xi $ for
different values of $\Gamma _j $ with $\lambda $=0.5.}\label{fig1}
\end{center}
\end{figure}

\quad\quad According to the conservation law of mass, the accretion
rate of disk matter is related to the mass outflow rate by

\begin{equation}
\label{eq14}
{d\dot {M}_{acc} (r)} \mathord{\left/ {\vphantom {{d\dot {M}_{acc} (r)}
{dr}}} \right. \kern-\nulldelimiterspace} {dr} = 4\pi r\dot {M}_{jet} .
\end{equation}

Incorporating equations (\ref{eq13}) and (\ref{eq14}), we have the
dimensionless accretion rate as follows,

\begin{equation}
\label{eq15}
\dot {m}_{acc} (\lambda ,\Gamma _j ;\xi ) = {\dot {M}_{acc} }
\mathord{\left/ {\vphantom {{\dot {M}_{acc} } {\dot {M}_H }}} \right.
\kern-\nulldelimiterspace} {\dot {M}_H } = 1 + \int_{\xi _H }^\xi {{\xi
}'\dot {m}_{jet} (\lambda ,\Gamma _j ,{\xi }')} d{\xi }'.
\end{equation}

\quad\quad By using equation (\ref{eq15}) we have the curves of
$\dot {m}_{acc} $ varying with $\xi $ for the given values of
$\Gamma _j $ as shown in Figure 2. It is shown that $\dot {m}_{acc}
$ increases steeply with the increasing $\xi $ in the innermost
ADAF, while it almost keeps constant as $r$ is greater than dozens
of $r_g $.

%-----------------------fig 2 ----------------------------------------

\begin{figure}
\vspace{0.5cm}
\begin{center}
\includegraphics[width=7cm]{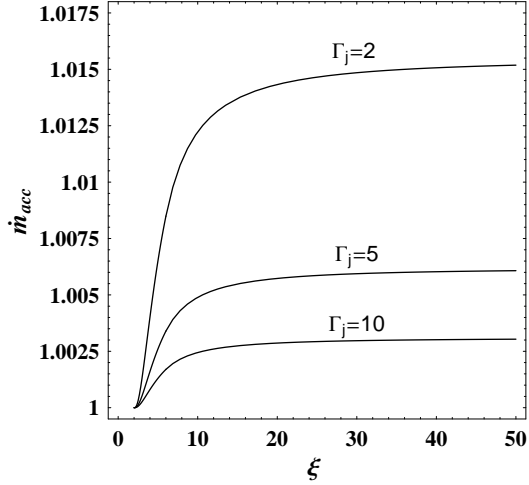}
\caption{ The curves of $\dot {m}_{acc} $ versus $\xi $ for
different values of $\Gamma _j $ with $\lambda $=0.5.}\label{fig2}
\end{center}
\end{figure}

\quad\quad In the BP process accreting matter is channeled into the
outflow/jet by virtue of the poloidal magnetic field lines frozen in
the disk, and the streaming gas is accelerated due to the work done
by the magnetic torque. Thus equation of angular momentum at the
presence of the jet should be written as

\begin{equation}
\label{eq16}
\frac{d}{dr}\left( {\dot {M}_{acc} \Omega r^2} \right) = -
\frac{d}{dr}\left( {4\pi r^2\tau _{r\varphi } H} \right) + 4\pi r\left[
{\dot {M}_{jet} \Omega r^2 + S_L } \right],
\end{equation}

\noindent
where $H$and $\tau _{r\varphi } $ are respectively the vertical scale height
and r$\phi $-component of stress tensor, and they read (Manmoto, Mineshige
{\&} Kusunose 1997)

\begin{equation}
\label{eq17}
H = {c_s } \mathord{\left/ {\vphantom {{c_s } {\Omega _K }}} \right.
\kern-\nulldelimiterspace} {\Omega _K } = {\sqrt {P \mathord{\left/
{\vphantom {P \rho }} \right. \kern-\nulldelimiterspace} \rho } }
\mathord{\left/ {\vphantom {{\sqrt {P \mathord{\left/ {\vphantom {P \rho }}
\right. \kern-\nulldelimiterspace} \rho } } {\Omega _K }}} \right.
\kern-\nulldelimiterspace} {\Omega _K },\mbox{ }\tau _{r\varphi } = - \alpha
P,
\end{equation}

\noindent
where $c_s , \quad P$ and $\rho $ are the isothermal sound speed, total pressure
and height-averaged density, respectively. Substituting equations (\ref{eq14}) and
(\ref{eq17}) into equation (\ref{eq16}), we have

\begin{equation}
\label{eq18}
\sqrt {\frac{P^3}{\rho }} = \frac{\sqrt {r_g c^2} }{4\pi \alpha
r^{3.5}}\left[ {\int_{r_H }^r {\dot {M}_{acc} (r)\frac{d}{dr}\left( {\Omega
r^2} \right)dr - \int_{r_{in} }^r {4\pi rS_L dr} } } \right],
\end{equation}

\noindent
where `no torque boundary condition' at the horizon of a Schwarzschild BH is
used in deriving equation (\ref{eq18}).

\quad\quad As given by Narayan {\&} Yi (1994), the parameter $\delta
$ indicates advection-dominated degree of the flow, and we have the
disk luminosity by integrating the energy equation as follows,

\begin{equation}
\label{eq19}
L_{disk} = \left( {1 - \delta } \right)\int_{r_H }^{r_{out} } {\tau
_{r\varphi } r\left( {{d\Omega } \mathord{\left/ {\vphantom {{d\Omega }
{dr}}} \right. \kern-\nulldelimiterspace} {dr}} \right)} 4\pi rHdr = \left(
{1 - \delta } \right)\int_{r_H }^{r_{out} } {6\pi \alpha fr\sqrt {{P^3}
\mathord{\left/ {\vphantom {{P^3} \rho }} \right. \kern-\nulldelimiterspace}
\rho } } dr.
\end{equation}

\quad\quad Substituting equation (\ref{eq18}) into equation
(\ref{eq19}), we express the disk luminosity of ADAF as

\begin{equation}
\label{eq20}
\begin{array}{l}
L_{disk} = 1.5\times \sqrt {r_g c^2} \left( {1 - \delta }
\right)\\
 \ \ \ \ \ \ \ \ \ \times \int_{r_H }^{r_{out} } {\frac{f}{r^{2.5}}\left[ {\int_{r_H
}^r {\dot {M}_{acc} ({r}')\frac{d}{d{r}'}\left( {\Omega {r}'^2}
\right)d{r}'} - \int_{r_H }^r {4\pi {r}'S_L d{r}'} } \right]} dr,\\
 \end{array}
\end{equation}

\noindent
where the first term in the bracket at the right-hand side represents the
releasing rate of the accreting matter's energy, and the second integral is
the cooling rate due to the outflow/jet driven by the BP process.

\quad\quad Substituting equations (\ref{eq1}), (\ref{eq3}),
(\ref{eq5}), (\ref{eq7}), (\ref{eq11}) and (\ref{eq18}) into
equation (\ref{eq20}), we express the dimensionless disk luminosity
as follows,

\begin{equation}
\label{eq21}
\begin{array}{l}
 l_{disk} = {L_{disk} } \mathord{\left/ {\vphantom {{L_{disk} } {\dot
{M}_{Edd} c^2}}} \right. \kern-\nulldelimiterspace} {\dot {M}_{Edd}
c^2} = 1.5\times \dot {m}_H ( {1 - \delta } )\\
 \ \ \ \times \int_{\xi _H
}^{\xi _{out} } {\frac{f}{\xi ^{2.5}}} \left[ {\int_{\xi _H }^\xi
{\dot {m}_{acc} (\lambda ,\Gamma _j ,{\xi }')\frac{d}{d{\xi
}'}\left( {f{\xi }'^{0.5}} \right)d{\xi }' - \frac{1}{\lambda
}\int_{\xi _H }^\xi {2\xi _H^{0.5} f{\xi }'^{ - 1}d{\xi }'}
} } \right]d\xi. \\
 \end{array}
\end{equation}

\quad\quad Four parameters, $\Gamma _j $, $\dot {m}_H $, $\lambda $
and $\delta $ are involved in equation (\ref{eq21}) for the disk
luminosity $l_{disk} $. In fact $l_{disk} $ is determined mainly by
the parameters $\dot {m}_H $, $\lambda $ and $\delta $, being
insensitive to the Lorentz factor $\Gamma _j $ as shown in Figure 3.
According to the observations of AGNs given by some authors (Urry
{\&} Padovani 1995; Biretta, Sparks {\&} Macchetto 1999), we take
the typical value, $\Gamma _j = 10$, for calculating jets in this
paper.

%-----------------------fig 3 ----------------------------------------

\begin{figure}
\vspace{0.5cm}
\begin{center}
\includegraphics[width=7cm]{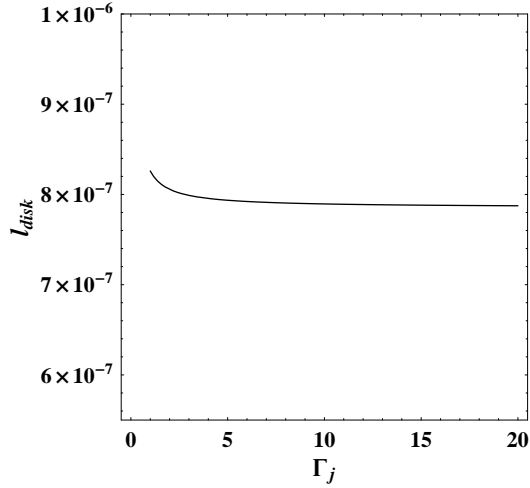}
\caption{ The curves of $l_{disk} $ versus $\Gamma _j $ with
$\lambda $=0.5 for $\dot {m}_H = 0.01$, $\delta = 0.99$ and $\xi
_{out} = 10^5$.}\label{fig3}
\end{center}
\end{figure}

\quad\quad By using equations (\ref{eq9}) and (\ref{eq11}) the jet
power driven by the BP process is expressed as

\begin{equation}
\label{eq22}
Q_{jet} = \int_{r_H }^{r_{out} } {\Gamma _j \dot {M}_{jet} c^2} 4\pi rdr =
\dot {M}_H r_H^{0.5} \lambda ^{ - 1}\int_{r_H }^{r_{out} } {r^{0.5}\Omega
^2} dr.
\end{equation}

\quad\quad Inspecting equation (\ref{eq22}), we find that the jet
power depends on the accretion rate at the innermost ADAF, the
parameter $\lambda $ and the angular velocity of the accretion flow.
Since $\dot {M}_H $ and $\lambda $ expressed by equation (\ref{eq5})
is related to the magnetic field strength at the horizon, and
$\Omega $ is linked to accretion mode, the jet power is expected to
be related to these two factors.

\quad\quad Substituting equations (\ref{eq5}), (\ref{eq11}) and
(\ref{eq12}) into equation (\ref{eq22}), we have the dimensionless
jet power expressed by

\begin{equation}
\label{eq23}
\begin{array}{l}
q_{jet} = {Q_{jet} } \mathord{\left/ {\vphantom {{Q_{jet} } {\dot
{M}_{Edd} c^2}}} \right. \kern-\nulldelimiterspace} {\dot {M}_{Edd}
c^2} \\
\mbox{ }\mbox{ }\mbox{ }\mbox{ }= 2\dot {m}_H f_0^2 \xi _H^{0.5}
\lambda ^{ - 1}\left[ {\int_{\xi _H }^6 {\xi ^{ -
2.5}\frac{9}{4}\left( {\frac{\xi - 2}{\xi }} \right)^2} d\xi +
\int_6^{\xi _{out} } {\xi
^{ - 2.5}} d\xi } \right].\\
\end{array}
\end{equation}

%-----------------------sec 3 ----------------------------------------

\section{FITTING JET POWERS AND X-RAY LUMINOSITIES OF A SAMPLE OF AGNS}

It is shown in MH07 that a clear correlation exists between Eddington-scaled
kinetic power and bolometric luminosity of 15 sub-Eddington AGNs. The
measured slope suggests that these objects are in a radiatively inefficient
accretion mode. Based on Kuncic {\&} Bicknell (2004) a simple coupled
accretion-outflow disk model was presented to explain the main features of
the observed sample. However, dynamically important magnetic fields and
their role in the jet production mechanisms were not taken into account in
MH07. We intend to fit the 2--10 keV luminosity and kinetic power of the 15
samples based on our model.

\quad\quad Inspecting equation (\ref{eq21}), we find that the
radiation flux at the presence of the jet could become negative for
$\lambda $ less than a critical value, and we obtain $\lambda _{\min
} = 0.447$ by setting the minimum radiation flux equal to zero.

\quad\quad Narayan {\&} Yi (1995) suggested that there is a
theoretical upper limit on the accretion rate for an ADAF. The
optically thin ADAF does not exist, and it transits to an optically
thick disk for the accretion rate greater than a critical one. The
exact value of the critical accretion rate is still unclear,
depending on the viscosity parameter $\alpha $, i.e. $\dot
{m}_{crit} \simeq 0.28\alpha ^2$ (Mahadevan 1997). In our model we
take $\dot {m}_{crit} \simeq 0.0252$ with $\alpha = 0.3$. Combining
equation (\ref{eq23}) with the relation $\dot {m}_H \le \dot
{m}_{crit} $, we have each maximum $\lambda _{\max } $ in fitting
the jet power of each source.

\quad\quad According to MH07, for all objects in the sample, the BH
mass $M_{BH} $ could be estimated either through the $M - \sigma $
relation or via direct dynamical measurements. As a simple analysis,
the disk luminosity is related to the 2--10 keV luminosity by
$L_{disk} = 5L_{2 - 10keV} $, and the jet power is equal to the
kinetic power, $Q_{jet} = L_{kin} $. With $\lambda _{\min } = 0.447$
and $\lambda _{\max } $ for each source we derive the 2--10 keV
luminosity and kinetic power of the 15 samples based on equations
(\ref{eq21}) and (\ref{eq23}) as shown in Table 1.

\begin{table*}
\caption{\label{tab1} The values of the concerned parameters for
fitting the 2--10 keV luminosity and kinetic power of the 15
sub-Eddington AGNs with $\lambda _{\min } $= 0.447 and $\lambda
_{\max } $, where $f_0 = 0.33$, $\alpha = 0.3$, $\Gamma _j = 10$ and
$\xi _{out} = 10^4$ are assumed.}

\begin{center}

\begin{tabular}{c c c c c c c}
 \hline  \hline \raisebox{-1.50ex}[0cm][0cm] {Object} &
 \raisebox{-1.50ex}[0cm][0cm] {$\lg M_{BH}$} &
 \raisebox{-1.50ex}[0cm][0cm] {$\lg L_{2-10keV}$} &
 \raisebox{-1.50ex}[0cm][0cm] {$\lg L_{Kin} $ }&
 \multicolumn{2}{c} {$\lambda_{\min}$=0.447} &
 $\lambda_{\max}$ \\
\cline{5-7}
 &
 &
 &
 &
$\lg \dot {m}_H $& $\lg (1 - \delta )$&
$\lg (1 - \delta )$ \\
 \hline Cyg A \quad & 9.40& 44.22& $45.41_{-0.1}^{+0.19} $& $
-1.63_{-0.1}^{+0.19} $& $0.51_{+0.1}^{-0.19} $&
$0.21_{+0.1}^{-0.19} $ \\
\hline NGC 507& 8.90& $<39.90$ & $44.01_{-0.26}^{+0.16} $&
$-2.53_{-0.26}^{+0.16}$ & $<-2.41_{+0.26}^{-0.16}$ &
$ <-4.41_{+0.26}^{-0.16} $ \\
\hline NGC 1275& 8.64& 43.40& $44.33_{-0.14}^{+0.17} $&
$-1.95_{-0.14}^{+0.17} $& $0.77_{+0.14}^{-0.17} $&
$-0.47_{+0.14}^{-0.17} $ \\
\hline NGC 4374& 8.80& 40.34& $42.59_{-0.5}^{+0.6} $& $
-3.85_{-0.5}^{+0.6} $& $ -0.55_{+0.5}^{-0.6} $&
$ -3.92_{+0.5}^{-0.6} $ \\
\hline NGC 4472& 8.90& 38.46& $42.91_{-0.23}^{+0.14} $& $
-3.63_{-0.23}^{+0.14} $& $ -2.75_{+0.23}^{-0.14} $&
$ -5.90_{+0.23}^{-0.14} $ \\
\hline NGC 4486& 9.48& 40.55& $43.44_{-0.5}^{+0.5} $&
$-3.68_{-0.5}^{+0.5} $& $ -1.19_{+0.5}^{-0.5} $&
$-4.39_{+0.5}^{-0.5} $ \\
\hline NGC 4552& 8.57& 39.33& $42.20_{-0.21}^{+0.14} $& $
-4.01_{-0.21}^{+0.14} $& $ -1.17_{+0.21}^{-0.14} $&
$-4.70_{+0.21}^{-0.14} $ \\
\hline NGC 4636& 8.20& $<38.40$ & $42.65_{-0.15}^{+0.11} $& $
-3.19_{-0.15}^{+0.11} $& $ <-2.55_{+0.15}^{-0.11} $&
$ <-0.63_{+0.15}^{-0.11} $ \\
\hline NGC 4696& 8.60& 40.26& $42.89_{-0.22}^{+0.22} $& $
-3.35_{-0.22}^{+0.22} $& $ -0.93_{+0.22}^{-0.22} $&
$ -3.79_{+0.22}^{-0.22} $ \\
\hline NGC 5846& 8.59& 38.37& $41.86_{-0.29}^{+0.18} $& $
-4.37_{-0.29}^{+0.18} $& $ -1.79_{+0.29}^{-0.18} $&
$ -4.83_{+0.29}^{-0.18} $ \\
\hline NGC 6166& 8.92& 40.56& $43.82_{-0.4}^{+0.5} $& $
-2.74_{-0.4}^{+0.5} $& $ -1.56_{+0.4}^{-0.5} $&
$ -3.79_{+0.4}^{-0.5} $ \\
\hline IC 4374& 8.57& 41.37& $43.30_{-0.26}^{+0.36} $& $
-2.91_{-0.26}^{+0.36} $& $ -0.23_{+0.26}^{-0.36} $&
$ -0.62_{+0.26}^{-0.36} $ \\
\hline UGC 9799& 8.58& 41.89& $44.18_{-0.28}^{+0.36} $& $
-2.04_{-0.28}^{+0.36} $& $ -0.59_{+0.28}^{-0.36} $&
$ -1.97_{+0.28}^{-0.36} $ \\
\hline 3C 218& 8.96& 42.17& $44.63_{-0.10}^{+0.16} $& $ -
1.97_{-0.10}^{+0.16} $& $ -0.76_{+0.10}^{-0.16} $&
$ -2.03_{+0.10}^{-0.16} $ \\
\hline 3C 388& 9.18& 41.69& $44.30_{-0.30}^{+0.38} $& $
-2.52_{-0.30}^{+0.38} $& $ -0.91_{+0.30}^{-0.38} $&
$ -2.90_{+0.30}^{-0.38} $\\
\hline \hline
\end{tabular}
\end{center}

\textbf{Notes: }Column (1): source name; column (2): logarithm of
the BH mass as derived from $M - \sigma $ relation; column (3):
logarithm of the intrinsic rest-frame luminosity in the 2--10 keV
band; column (4): logarithm of the kinetic luminosity; column (5):
logarithm of the dimensionless accretion rate at the BH horizon with
the lower limit of $\lambda _{min} = 0.447$; column (6): logarithm
of $1 - \delta $ with $\lambda _{\min } = 0.447$; column (7):
logarithm of $1 - \delta $ with $\lambda _{\max } $ and $\dot {m}_H
= \dot {m}_{crit} $.

\end{table*}

%-----------------------fig 4 ----------------------------------------

\begin{figure}
\vspace{0.5cm}
\begin{center}
\includegraphics[width=7cm]{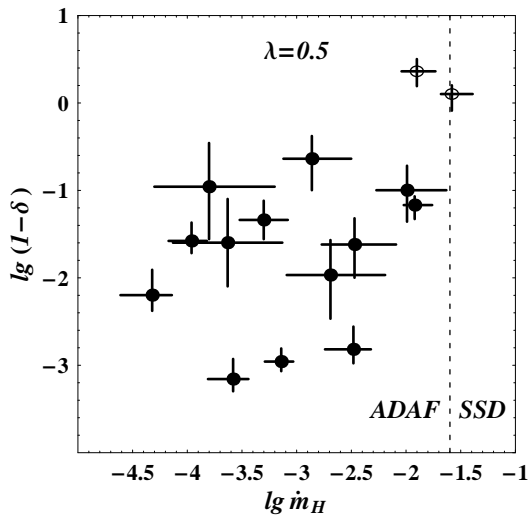}
\caption{ The $\lg \dot {m}_H - \lg (1 - \delta )$ parameter space:
the values of $\lg \dot {m}_H $ and $\lg (1 - \delta )$ for fitting
the 2--10 keV luminosity and kinetic powers of the 15 sub-Eddington
AGNs with $\lambda = 0.5$. The dashed line represents the critical
accretion rate for ADAFs.}\label{fig4}
\end{center}
\end{figure}

\quad\quad From Table 1 and Figure 4 we find that most of the
samples are consistent with the ADAF model except Cyg A and NGC
1275, which are marked by the open circles in Figure 4 with $\lambda
= 0.5$. These two sources of negative advection parameter $\delta $
are conflict with the ADAF model. As shown in column (\ref{eq7}) of
Table 1, the value of $\delta $ could be positive for NGC 1275,
provided that $\lambda $ is not so small. However, the negative
$\delta $ cannot be removed for Cyg A.

\quad\quad From Table 1 and Figure 4 we find that the higher
dimensionless accretion rate is apt to correspond to the greater
$\lg \left( {1 - \delta } \right)$, implying an anti-correlation of
the accretion rate $\dot {m}_H $ with the advection parameter
$\delta $. This result is consistent with the fact that there is a
theoretical upper limit on the accretion rate for an ADAF: An
optically thin ADAF transits to an optically thick disk as the
accretion rate reaches the critical one. This anti-correlation of
accretion rate with the advection parameter $\delta $ could be used
to explain the correlation between Eddington-scaled kinetic power
and bolometric luminosity of the sub-Eddington AGNs.

%-----------------------sec 4 ----------------------------------------

\section{THE LEDLOW-OWEN RELATION FOR FR I/II DICHOTOMY}

Very recently, WC08 reproduce the dividing line of the Ledlow-Owen relation
for FR I/II dichotomy, being related to the jet power and BH mass as
follows,

\begin{equation}
\label{eq24}
\lg Q_{jet} (erg\mbox{ }s^{ - 1}) = 1.13\lg M_{BH} (M_ \odot ) + 33.42 +
1.50\lg F.
\end{equation}

\quad\quad In equation (\ref{eq24}) the factor $F$ parameterizes the
uncertainties of the normalization, which is constrained to be
between 10 and 20 (Blundell {\&} Rawlings 2000).

Replacing $Q_{jet} $ in equation (\ref{eq24}) by accretion rate given by equation
(\ref{eq23}), we have another way to express the Ledlow-Owen relation as follows,

\begin{equation}
\label{eq25}
\begin{array}{l}
 \lg \dot {m}_H = 0.13\lg M_{BH} (M_ \odot ) - 5.68 + 1.50\lg F \\
 \ \ \ \ \ \ \ \ \ \ \ - \lg \left\{ {2f_0^2 \xi _H^{0.5} \lambda ^{ - 1}\left[
{\int_{\xi _H }^6 {\xi ^{ - 2.5}\frac{9}{4}\left( {\frac{\xi -
2}{\xi }} \right)^2} d\xi + \int_6^{\xi _{out} } {\xi ^{ - 2.5}}
d\xi } \right]}
\right\} \\
 \end{array}
\end{equation}

\quad\quad There is growing evidence suggesting that most FR I type
radio galaxy nuclei may possess ADAFs (Reynolds et al. 1996; Gliozzi
et al. 2003; Merloni et al. 2003; Donato et al. 2004; Wu et al.
2007). This implies that the FR I/II dividing dimensionless
accretion rate should be equal to or less than the upper limit of
the accretion rate for ADAF. Incorporating equation (\ref{eq25}) and
the relation $\dot {m}_H = \dot {m}_{crit} $ we can derive an upper
limit, $\lambda \le 1.55$.

\quad\quad Based on equation (\ref{eq25}) we have the dividing lines
for FR I/II dichotomy in the plane of $\lg M_{BH} $ versus $\lg \dot
{m}_H $ with $\lambda = $0.447 and 1.55 as shown in Figures 5a and
5b, respectively. It is found that the FR I/II dichotomy is
determined mainly by accretion rate at the BH horizon, being
insensitive to the BH mass. This result implies that FR I radio
sources correspond to lower $\dot {m}_H $, whereas FR II radio
galaxies have larger $\dot {m}_H $.

\quad\quad As shown in Figure 5, the FR I/II division can be
determined by a constant $\dot {m}_H $ with the given values of the
parameters $F$ and $\lambda $. It is noted that all these dividing
lines are below the dotted lines for $\lg \dot {m}_{crit} $, and
this result strongly suggests that the FR I sources are in the ADAF
state.

\quad\quad A weak BH mass dependence and a separation luminosity of
about $10^{ - 3} - 10^{ - 2}$ Eddington accretion rate in the FR
I/II dichotomy have been found in GC01 and WLA07, and our model are
consistent with these results.

%-----------------------fig 5 ----------------------------------------

\begin{figure}
\vspace{0.5cm}
\begin{center}
{\includegraphics[width=6cm]{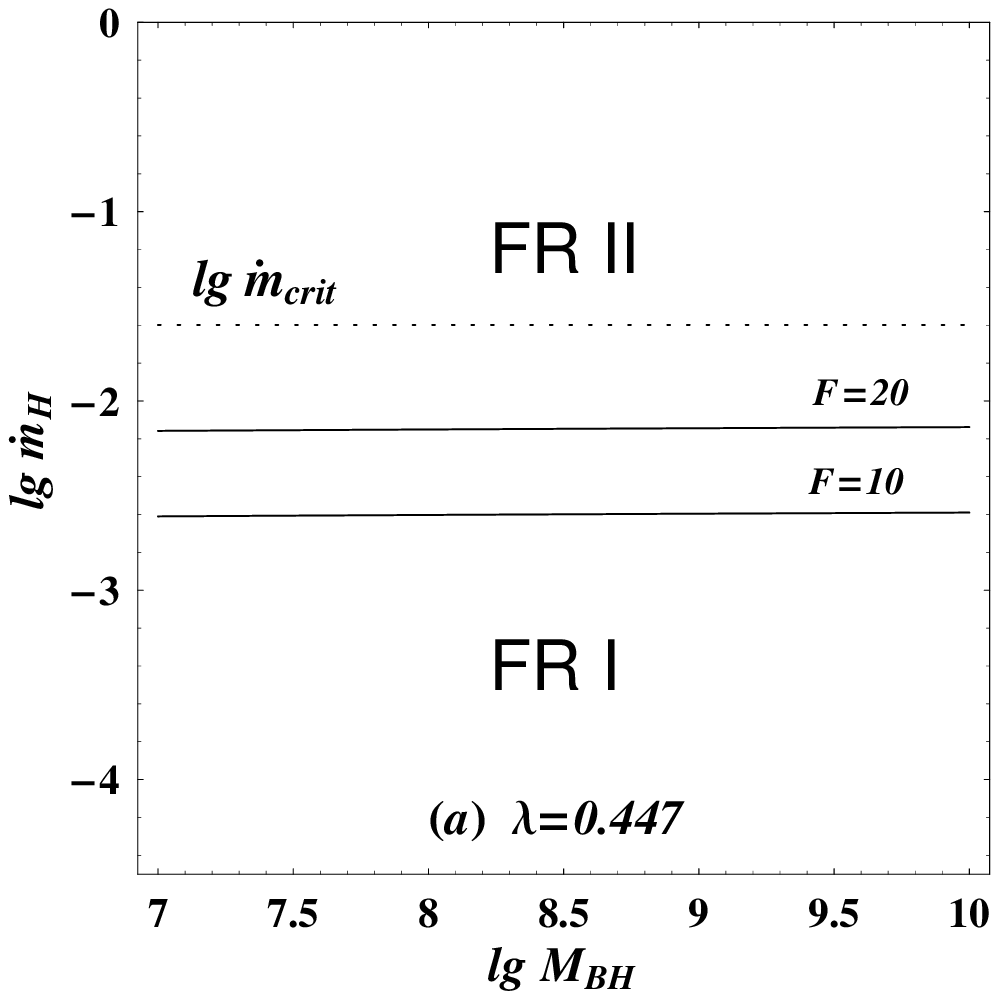}
 \includegraphics[width=6cm]{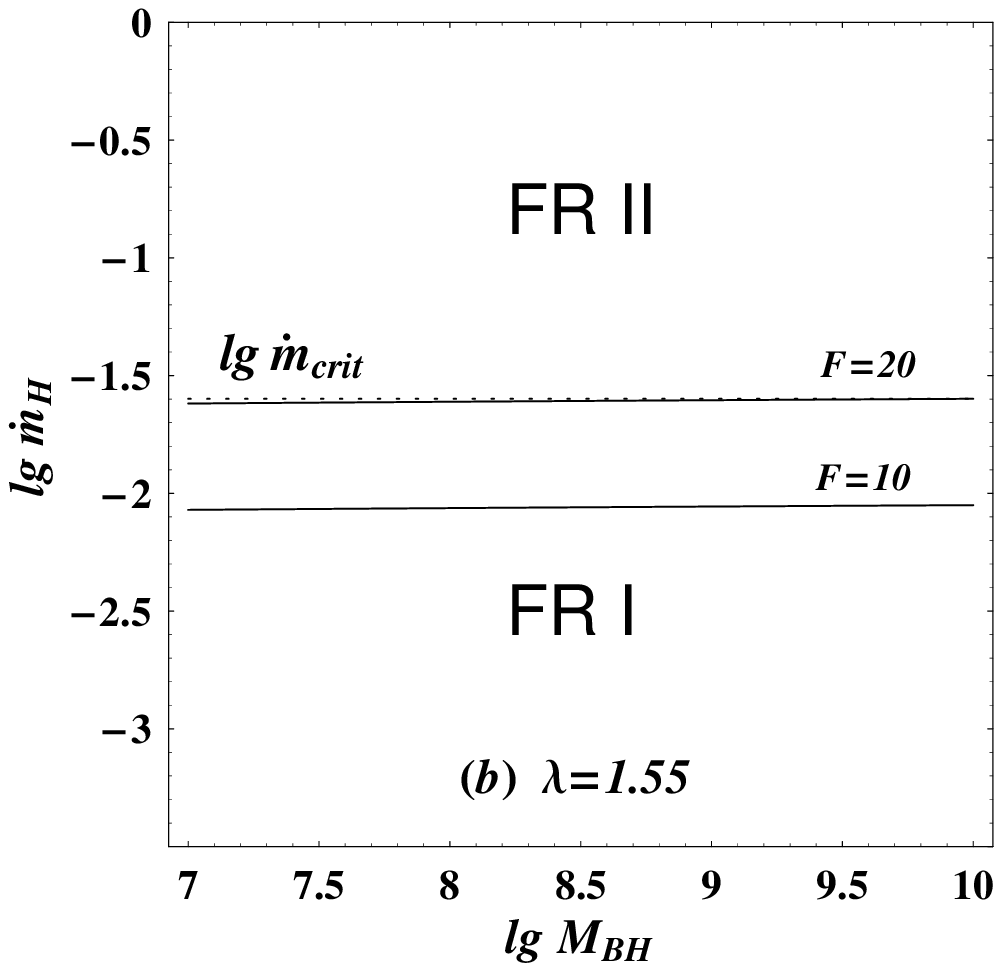}}
\caption{ The dividing lines for FR I/II dichotomy (solid lines) in
$\lg M_{BH} - \lg \dot {m}_H $ plane with $\lambda = $0.447 and 1.55
in left and right panels, respectively. The dotted line represents
the critical accretion rate for ADAFs.}\label{fig5}
\end{center}
\end{figure}

%-----------------------sec 5 ----------------------------------------

\section{SUMMARY}

In this paper, we discuss the outflow/jet driven by the BP process, and
investigate the interaction of the outflow/jet with ADAF. Based on the
conservation laws of mass, angular momentum and energy we derive the
expressions of the BP power and disk luminosity. The disk luminosity is
suppressed significantly due to two reasons, (\ref{eq1}) a fraction of accretion
energy is channeled into the outflow/jet via the BP process, and (\ref{eq2}) most of
the accretion energy remaining in the gas is advected into the BH in the
form of entropy.

\quad\quad Based on our model we fit the 2--10 keV luminosity and
kinetic power of the 15 samples, and mark the points of these
sources in the $\lg \dot {m}_H - \lg (1 - \delta )$ parameter space.
It is found that there exists an anti-correlation between the
accretion rate and the advection parameter, which could be used to
explain the correlation between Eddington-scaled kinetic power and
bolometric luminosity of the 15 samples.

\quad\quad We find that the parameter $\lambda $ is very important
in adjusting the magnetic pressure at the BH horizon relative to the
ram pressure of ADAF, and its value range can be determined by
combining some theoretical consideration with the observations as
argued in Sect. 3 and Sect. 4.

\quad\quad As a simplified model, only a Schwarzschild BH is
involved, and we fail to consider the BZ process and the effects of
BH spins on jet powers and disk luminosities of the AGNs. We shall
improve the model in our future work.

\textbf{Acknowledgments.} This work is supported by the National Natural
Science Foundation of China under grant 10873005, the Research Fund for the
Doctoral Program of Higher Education under grant 200804870050 and National
Basic Research Program of China -- 973 Program 2009CB824800. We are very
grateful to the anonymous referee for his (her) instructive comments on the
manuscript.

{}

\begin{appendix}
\section{\textbf{APPENDIX A: DERIVATION OF EQUATION (6)}}

Following Wang et al. (2004, hereafter W04), we express the Poynting
energy flux $S_E $ as
\begin{equation}
\label{A2}
 S_E = \frac{\Delta P_{EM} }{2\pi r\Delta r}
\end{equation}

\noindent where $\Delta P_{EM} $ is the electromagnetic power
extracted from ADAF between two adjacent magnetic surfaces. By
invoking an equivalent circuit given by W04 $\Delta P_{EM} $ is
written as

\begin{equation}
\label{A2}
 \Delta P_{EM} = \left( {{\Delta \varepsilon } \mathord{\left/ {\vphantom
{{\Delta \varepsilon } {\Delta Z_A }}} \right.
\kern-\nulldelimiterspace} {\Delta Z_A }} \right)^2\Delta Z_A ,
\end{equation}

\noindent where $\Delta \varepsilon $ is the electromotive force in
the equivalent circuit, and it reads

\begin{equation}
\label{A3}
 \Delta \varepsilon = - \left( {{\Delta \Psi } \mathord{\left/ {\vphantom
{{\Delta \Psi } {2\pi }}} \right. \kern-\nulldelimiterspace} {2\pi
}} \right)\Omega ,
\end{equation}

\noindent and $\Delta \Psi = B_{ADAF} 2\pi rdr$ is the magnetic flux
between the two adjacent magnetic surfaces. In W04 the load
resistance is assumed to be axisymmetric, being located evenly in a
plane,

\noindent and $\Delta Z_A $ is the resistance between the two
adjacent magnetic surfaces. The surface resistivity $\sigma _L $ of
the unknown load is assumed to obey the following relation,

\begin{equation}
\label{A4} \sigma _L = \alpha _Z \sigma _H = 4\pi \alpha _Z ,
\end{equation}

\noindent where $\alpha _Z $ is a parameter, and $\sigma _H = 4\pi =
377\mbox{ }ohm$ is the surface resistivity of the BH horizon.

\begin{equation}
\label{A5} \Delta Z_A = 2\alpha _Z \frac{\Delta{r}}{r} ,
\end{equation}

Assuming $\alpha _Z = 1$and incorporating equation (A.1)---(A.5), we
derive equation (\ref{eq6}) for the Poynting energy flux $S_E $.

\end{appendix}

\end{document}